\documentclass[aps,twocolumn,tightenlines,nofootinbib,superscriptaddress]{revtex4-2}

\usepackage[utf8]{inputenc}
\usepackage{graphicx}
\usepackage{mwe}
\usepackage{mathrsfs}
\usepackage{amsfonts}
\usepackage{setspace}
\usepackage{cellspace}
\usepackage{amsmath,amssymb,bm}
\usepackage[colorlinks=true,linkcolor=blue]{hyperref}
\usepackage{xcolor}
\usepackage{epsfig}
\usepackage{slashed}
\usepackage{hhline,multirow,tabularx}  %
\usepackage{dcolumn}    %
\usepackage{url}        %
\usepackage{braket}     %
\usepackage{makecell}
\usepackage{bbm}

\newcommand{\bfb}{\mathbf{b}}

\newcommand{\bfx}{\mathbf{x}}
\begin{document}

\title{From wave-function to fireball geometry: the role of a restored broken symmetry in ultra-relativistic collisions of deformed nuclei}

\author{Yanru Bao}
\email{yanrubao@mails.ccnu.edu.cn}
\affiliation{Key Laboratory of Quark and Lepton Physics (MOE) \& Institute of Particle Physics, Central China Normal University, Wuhan, Hubei 430079, China}

\author{Weiyao Ke}
\email{weiyaoke@ccnu.edu.cn}
\affiliation{Key Laboratory of Quark and Lepton Physics (MOE) \& Institute of Particle Physics, Central China Normal University, Wuhan, Hubei 430079, China}
\affiliation{Southern Center for Nuclear-Science Theory (SCNT), Institute of Modern Physics, Chinese Academy of Sciences, Huizhou, Guangdong 516000, China}
\date{\today}

\begin{abstract}
Monte Carlo (MC) Glauber models for high-energy nuclear collisions conventionally treat deformed nuclei as classical rotors, encoding deformation in the one-body distribution of a single intrinsic state. The physical nuclear ground state, however, is rotationally invariant and describes a quantum rotor: a coherent superposition of intrinsic states at different orientations, with shape information encoded in many-nucleon correlations. For axially deformed even-even nuclei, we develop a modified MC Glauber sampling procedure consistent with this rotational symmetry. The model retains information off-diagonal in the basis of orientations through an angular overlap function and a generalized thickness function. We provide a practical prescription for computing both quantities and show that the rotational symmetry of the nuclear ground state is important in central collisions involving light, strongly deformed nuclei: for ${}^{20}\mathrm{Ne}$, the elliptic eccentricity is reduced by approximately $10\%$ level relative to the classical-rotor treatment. These results highlight the importance of incorporating quantum-mechanical constraints when relating intrinsic nuclear deformation to the QGP geometry in relativistic nuclear collisions.
\end{abstract}

\maketitle

{\bf \noindent Introduction.---} 
In ultra-relativistic heavy-ion collisions (HICs), inelastic scattering in the nuclear overlap region deposits sufficient energy and entropy to create a fireball of strongly-coupled quark-gluon plasma (QGP)~\cite{Shuryak:2014zxa}. The QGP fireball expands hydrodynamically, converting initial spatial eccentricities into final-state momentum anisotropies~\cite{Heinz:2013th,Heinz:2024jwu}. Measurements of these anisotropies and their correlations thus provide a window into the initial QGP geometry, which carries imprints of the structural information of the two colliding nuclei, such as their deformation.

This connection has driven extensive investigations of nuclear structure effects in HICs~\cite{Shou:2014eya,Goldschmidt:2015kpa,Li:2018oec,Giacalone:2019pca,Giacalone:2020awm,Jia:2021wbq,Jia:2021oyt,Giacalone:2021uhj,Giacalone:2021udy,Zhang:2022fou,Ma:2022dbh,Jia:2022ozr,Nie:2022gbg,Giacalone:2023hwk,Nie:2023dla,Zhang:2024bcb,Giacalone:2025vxa,Huang:2025cjm,Luzum:2023gwy,Blaizot:2025scr,Duguet:2025hwi,Giacalone:2025vxa}, with significant attention devoted to deformation effects in isobaric Zr-Zr/Ru-Ru collisions~\cite{Zhang:2021kxj,Wang:2024mce}, ${}^{238}$U~\cite{Ryssens:2023fkv,STAR:2024wgy,Xu:2024bdh,Xu:2025cgx,Zhang:2025hvi,Liu:2025fnq}, ${}^{197}$Au~\cite{Bally:2023dxi}, and Xe deformation and its shape phase transitions~\cite{Bally:2022rhf,Mantysaari:2024uwn,Zhao:2024lpc}. Beyond deformation, these studies have also targeted constraining neutron skin and symmetry energy~\cite{Li:2019kkh,Xu:2021qjw,Xu:2021vpn,Liu:2022xlm,Liu:2022kvz,Jia:2022qgl,Liu:2023pav,Giacalone:2023cet,Hao-Jie:2024nnh}, many-body correlations~\cite{Mehrabpour:2026yuc,Giacalone:2026fat}, and nuclear transition matrix elements~\cite{Li:2025vdp}.
Light nuclei collisions have also emerged as a particularly active frontier due to recent experiments at RHIC and the LHC~\cite{Brewer:2021kiv,Ding:2023ibq,Giacalone:2024luz,YuanyuanWang:2024sgp,Giacalone:2024ixe,Zhang:2024vkh,Zhao:2024feh,Hu:2025eid,Nijs:2025qxm,Liu:2025zsi,Huang:2023viw,ALICE:2025luc,ATLAS:2025nnt,CMS-PAS-HIN-25-009,LHCb:2022qvj,Smith:2025pol,Mariani:2942052}. Complementary probes of nuclear deformation are also being pursued in ultra-peripheral collisions and future electron-ion colliders, where diffractive processes access nuclear structure without forming a QGP~\cite{Mantysaari:2023qsq,Mantysaari:2023prg,Mantysaari:2023gop,Lin:2024mnj,Mantysaari:2024xmy}.

However, concerns have been raised regarding the treatment of nuclear deformation in HIC modeling~\cite{Dobaczewski:2025rdi}. Nuclear deformation is defined in the intrinsic frame, where a many-nucleon configuration $|\Omega, \beta_n\rangle$ is characterized by deformations $\beta_n$ and an orientation $\Omega$ relative to the lab frame.  For simplicity, this work only discuss even-even nuclei with axial deformation. Clearly, $|\Omega, \beta_n\rangle$ breaks SO(3) rotational symmetry, but a valid ground state must restore it. This can be achieved by applying the angular momentum projection operation $|JMK\rangle\propto \hat{P}_{MK}^J|\Omega, \beta_n\rangle$~\cite{Ring1980}, with
\begin{align}
\hat{P}_{MK}^J = \frac{2J+1}{8\pi^2}\int_{\alpha,\beta,\gamma} (D^J_{MK})^* (\alpha,\beta,\gamma) \hat{R}(\alpha,\beta,\gamma)\,.
\end{align}
$\hat{R}$ is the rotation operator with Euler angles ($\alpha, \beta,\gamma$), and $D^J_{MK}$ is the Wigner-$D$ matrix of representation $2J+1$. The projection introduces superposition of configurations at different orientations, and form a ``quantum rotor'' with good angular momentum quantum number $JMK$. Especially, $J=0$ state is rotational invariant and its one-body density is strictly spherical. So, deformation must be understood in terms of nucleon correlations. 
In contrast, traditional Monte Carlo (MC) Glauber models in relativistic HICs~\cite{Miller:2007ri,Shou:2014eya,Loizides:2025ule} treat deformed nuclei as classical rotors, encoding deformation solely through the one-body density of symmetry-breaking configurations with random orientations assigned in each collision event, which is incompatible with quantum mechanics.

Because MC Glauber models provide the participant nucleons' positions that serve as input to energy deposition models and thereby determine the initial QGP fireball geometry, we must investigate how deformation can be consistently incorporated within a quantum rotor framework. We propose to extend the conventional MC Glauber model with a modified nucleon position sampling procedure that approximately accounts for symmetry restoration. 

We remark that there are ``complete Glauber calculation'' which uses full nuclear wave functions with the correct symmetry~\cite{Horiuchi:2012ca,Hatakeyama:2019jfg}, but the studies focus on nuclear cross sections rather than QGP geometry. Finally, throughout the Letter, we use natural units $\hbar=c=1$.

{\bf \noindent Shape as many-body correlations in HIC.---} 
We start by discussing the total inelastic cross section between nuclei $A$ and $B$ within the Glauber theory~\cite{Glauber:1970jm,Glauber:1955qq},
\begin{align}
\sigma_{AB}^{\rm inel} = \int d^2\mathbf{b} \, \left( 1-\left|\langle A, B |S(\mathbf{b}) |A,B\rangle\right|^2 \right),
\end{align}
where $\mathbf{b}$ is the impact parameter and $S(\mathbf{b})$ is the elastic $S$-matrix element. The integrand gives the probability that the initial nuclei do not remain in their ground states. At relativistic energies, the QGP is studied through particle-producing nuclear breakup events, which constitute the experimentally selected non-single-diffractive event sample~\cite{ALICE:2013hur}. We therefore exclude final states that are purely nuclear-level excitations $A^*$ and $B^*$, and compute the non-diffractive (ND) inelastic cross section (the difference between ND and NSD is negligible for our purpose, see Supplemental Material I),
\begin{align}
\sigma_{AB}^{\rm ND} =& \int d^2\mathbf{b} \, \Big( 1-\sum_{A^*,B^*}\left|\langle A^*, B^* |S(\mathbf{b}) |A,B\rangle\right|^2 \Big) \nonumber \\
=& \int d^2\mathbf{b} \, \left( 1-\big\langle  |S(\mathbf{b})|^2\big\rangle\right) 
\label{eq:ND-sigma}
\end{align}
The sums over $A^*$ and $B^*$ resolve the identity operators in the corresponding nuclear Hilbert spaces, leaving a ground-state expectation value of $|S(\bfb)|^2$. Here we adopt the shorthand notation $\langle\hat O\rangle\equiv\langle A,B|\hat O|A,B\rangle$.

In the Glauber theory~\cite{Glauber:1970jm,Glauber:1955qq}, the $S$ matrix of $AB$ collision is approximated by the product of binary-nucleon ($N$-$N$) collisions. Here, we only needs its squared form
\begin{align}
|S(\mathbf{b})|^2 &\approx \prod_{a=1}^A \prod_{b=1}^B\left(1-\mathcal{C}_{ab}\right)\,.
\label{eq:main_Glauber-S} 
\end{align}
$\mathcal{C}_{ab} = \mathcal{C}(\hat{\bm{x}}_a - \hat{\bm{x}}_b - \mathbf{b})$ is the probability of an inelastic collision between nucleons $a,b$, as a function of transverse separation.
The integrand of Eq.~\eqref{eq:ND-sigma} is interpreted as the probability of inelastic $AB$ interaction at a given $\bfb$. 
\begin{align}
P_{AB}^{\rm inel}(\mathbf{b}) &\equiv 1-\big\langle  |S(\mathbf{b})|^2\big\rangle \approx 1-\Big\langle\prod_{a,b}\big(1-\mathcal{C}_{ab}\big)\Big\rangle \label{eq:main_Glauber-P}\\
&= \sum_{(a,b)}\langle \mathcal{C}_{ab}\rangle - \frac{1}{2}\sum_{(a,b)\neq (a',b')}\langle \mathcal{C}_{ab}\mathcal{C}_{a'b'}\rangle +\cdots \,. \label{eq:main_Glauber-P2} 
\end{align}
So, evaluating Eq.~\eqref{eq:main_Glauber-P} requires expanding the products and computing expectation values of $1$-, $2$-, ..., $N$-body density matrices. Terms involving only a single interacting pair (peripheral collisions) require only the one-body density matrix and are spherically symmetric, carrying no deformation information. In the opposite limit of ultra-central collisions, all nucleons participate and the full $N$-body density matrices are required, retaining correlations such as deformation~\cite{Sun:2026yrr}. The nuclear shape thus emerges with a growing number of participants.

Alternatively, the MC Glauber model provides a probabilistic realization of Eq.~\eqref{eq:main_Glauber-P}: in each event, nucleon positions are sampled from the corresponding many-body spatial distributions, and participant nucleons are identified by sampling the $N$--$N$ collision probability $\mathcal{C}_{ab}$ for each nucleon pair. The initial QGP fireball geometry is then constructed from the energy deposited by these participant nucleons and fluctuates event by event.

In principle, if the full $N$-body spatial distribution can be calculated from first principles, nucleon configurations can be sampled directly without further approximations. Such a strategy has been implemented using nuclear lattice effective field theory (NLEFT) and its nucleon-position sampling technique~\cite{Lee:2025req,Giacalone:2024luz}. We instead addresses situations in which such many-body information is unavailable and develops an extension of the conventional MC sampling procedure that incorporates correlations induced by rotational symmetry restoration.

{\bf \noindent The modified MC Glauber sampling for quantum rotors---} 
We write down the $J=0$ ground state of an even-even nucleus ($X=A, B$) with axial deformation as
\begin{align}
|X\rangle
&=
\frac{1}{N_X}\int d\Omega_X\,|\Omega_X\rangle\,.
\label{eq:wavefunction}
\end{align}
It is a superposition of the deformed intrinsic state $|\Omega_X\rangle$ over different orientations $\Omega_X$, and we have suppressed writing $\beta_n$ explicitly. $N_X$ is the normalization with $|N_X|^2=\int d\Omega_X d\Omega_X' \langle \Omega_X|\Omega_X'\rangle$. For brevity, we also define a transitional expectation between intrinsic configurations with different orientations
\begin{align}
\overline{\mathcal{O}} = \frac{\langle\Omega_A, \Omega_B|
\hat{\mathcal{O}}|\Omega_A', \Omega_B'\rangle}{ \langle \Omega_A|\Omega_A'\rangle\langle \Omega_B|\Omega_B'\rangle}\,.
\end{align}
This is not to be confused with $\langle \hat{O} \rangle$, which is the expectation on the symmetry-restored ground state.

Plug Eq.~\eqref{eq:wavefunction} into the collision probability in Eq.~\eqref{eq:main_Glauber-P},  
\begin{align}
P_{AB}^{\rm inel}(\mathbf{b}) &= \int d\Omega_A \int d\Omega_A' \, \langle \Omega_A| \Omega_A'\rangle / |N_A|^2\nonumber\\
&\quad \times \int d\Omega_B \int d\Omega_B' \, \langle \Omega_B| \Omega_B'\rangle / |N_B|^2 \nonumber\\
&\quad \times \Big\{ 1-\overline{\prod_{a,b} \left[1-\mathcal{C}_{ab}(\mathbf{b})\right]}\Big\}\,.
\label{eq:New-Glauber-1}
\end{align}
The inner product $\langle \Omega_X| \Omega_X'\rangle$ is the overlapping function between intrinsic states at two different orientations. It is real for even-even ground state from time-reversal symmetry. We evaluate it using the Gaussian overlap approximation (GOA)~\cite{Ring1980} in a frame where $\Omega_X = \hat{z}$ and $\Omega_X'$ is connected by a rotation around the $y$-axis by $\theta$,
\begin{align}
\langle \Omega_X| \Omega_X'\rangle&=\langle \hat{z}|e^{-i\theta \hat{J}_y}|\hat{z}\rangle \nonumber\\
&\approx e^{-\frac{1}{2} \langle \hat{z}|\hat{J}_y^2|\hat{z}\rangle \sin^2\theta } \equiv \langle \Omega_X| \Omega_X'\rangle_{\rm GOA}\,.
\end{align}
The angular separation at which a deformed configuration and its rotated copy decorrelate is inversely proportional to the standard deviation of $\hat{J}_y$. Although $\langle \hat{z}|\hat{J}_y^2|\hat{z}\rangle$ is not an observable and depends on the many-body framework, qualitatively it increases with mass number and deformation. Therefore, decorrelation is significant for heavy nuclei or large deformations. Conversely, coherent superposition over orientations remains important for light nuclei or nuclei with small deformations.

To simplify the expectation value of the many-body collision product in Eq.~\eqref{eq:New-Glauber-1}, we adopt the independent-pair approximation (see Supplemental Material II), which resums all same-pair cumulants while neglects cumulants involving distinct binary collision pairs. Then,
\begin{align}
\overline{\prod_{a,b} \left[1-\mathcal{C}_{ab}(\mathbf{b})\right]} \approx  \left[1-
P_{\Omega_A\Omega_A';\Omega_B\Omega_B'}(\bfb)
\right]^{AB}\,,
\end{align}
where the binary collision probability is given by
\begin{align}
P_{\Omega_A\Omega_A';\Omega_B\Omega_B'}(\bfb) &= \int d^2\bfx_a\,d^2\bfx_b\,
\mathcal{C}\left(
\mathbf x_a-\mathbf x_b-\mathbf b
\right) \nonumber\\
& 
T_A
\left(
\mathbf x_a;\Omega_A,\Omega_A'
\right) T_B
\left(
\mathbf x_b;\Omega_B,\Omega_B'
\right)\,.
\label{eq:coll-pab}
\end{align}
$T_X$ is the generalized thickness functions, defined as
\begin{align}
T_X\left(\mathbf x;\Omega_X,\Omega_X'\right)\equiv\frac{1}{X}\int dz\,\frac{\langle\Omega_X|\hat{n}_X(\mathbf x,z)|\Omega_X'\rangle}{\langle\Omega_X|\Omega_X'\rangle}\,. \label{eq:TX}
\end{align}
$\hat{n}_X(\bfx, z)$ is the local density operator for nucleons in nucleus $X=A, V$. Note that it depends on two sets of orientations. On the diagonal $\Omega_X=\Omega_X'$, it is the same as the thickness function used in classical-rotor MC Glauber model, but $T_X$ also contains off-diagonal information.

Therefore, under the GOA and independent-pair approximations, Eq.~\eqref{eq:New-Glauber-1} becomes
\begin{align}
P_{AB}^{\rm inel}(\mathbf{b}) &\approx \int d\Omega_A \int d\Omega_A' \, \langle \Omega_A| \Omega_A'\rangle_{\rm GOA} / |N_A|_{\rm GOA}^2 \nonumber\\
&\quad \times \int d\Omega_B \int d\Omega_B' \, \langle \Omega_B| \Omega_B'\rangle_{\rm GOA} / |N_B|_{\rm GOA}^2 \nonumber\\
&\quad \times \Big\{ 1-\left[1-P_{\Omega_A\Omega_A';\Omega_B\Omega_B'}(\bfb)\right]^{AB}\Big\}\,. \label{eq:New-Glauber-2}
\end{align}
Because the GOA overlap functions are real and positive for even--even nuclei, while the collision term has the standard binomial form, Eq.~\eqref{eq:New-Glauber-2} admits a probabilistic interpretation. For each nucleus, we sample a pair of orientations $(\Omega_X,\Omega_X')$ according to the normalized de-correlation function $\langle \Omega_A| \Omega_A'\rangle_{\rm GOA}/|N_X|_{\rm GOA}^2$, and then sample nucleon transverse positions from the corresponding generalized thickness $T_X(\bfx;\Omega_X,\Omega_X')$. Binary collisions are subsequently sampled for each nucleon pair according to the profile $\mathcal C(\bfx_a-\bfx_b-\bfb)$. 

\begin{figure}[!t]
\centering
\includegraphics[scale=.67]{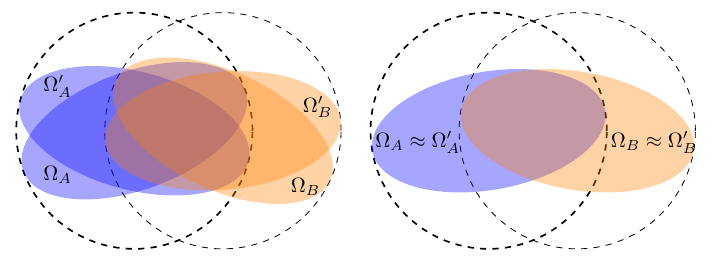}
\caption{Left: off-diagonal information in orientation is retained. Right: for sufficiently large nuclei, the de-correlation is strong, which recovers the classical-rotor limit.}
\label{fig:geo}
\end{figure}

Eqs.~\eqref{eq:coll-pab},~\eqref{eq:TX},~\eqref{eq:New-Glauber-2} and its probabilistic interpretation along with the modified sampling procedure constitutes the central result of this letter, and it deserves more discussion. The sampling depends on one set of transverse positions $\bfx$, but on two sets of orientations $\Omega_X, \Omega_X'$, as illustrated on the left of Fig.~\ref{fig:geo}. The dependence on both $\Omega_X$ and $\Omega_X'$ has a physical origin: in the eikonal limit, an inelastic $N$-$N$ interaction probes the transverse separation of the two nucleons. The resulting high-multiplicity final state acts as an environment~\cite{Zurek:2003zz}, and tracing over its unobserved degrees of freedom decoheres the nuclear wave function in the transverse-position basis in a short time scale~\cite{Muller:2005yu,Fries:2008vp}. This justifies the event-by-event sampling of nucleon transverse positions at an early stage of the collision and generates the fluctuations relevant to flow harmonics and correlation observables~\cite{Alver:2010gr,Heinz:2013th,Horecny:2025mak}. This decoherence, however, does not project the generalized coordinates $(\Omega,\beta_n,\ldots)$ that label the intrinsic configurations. The bra and ket may therefore retain different orientations, $\Omega_X$ and $\Omega_X'$, so the theory must depends on off-diagonal matrix elements in the generalized-coordinate basis (see Supplemental Materials IV.). This dependence is encoded by the generalized thickness function $T_X(\bfx;\Omega_X,\Omega_X')$ and the overlap function $\langle\Omega_X|\Omega_X' \rangle$. 
Therefore, although the independent-pair approximation neglects intrinsic nucleon correlations within each configuration, the effects of symmetry restoration are retained through the coherent superposition of differently oriented intrinsic states and are encoded in the overlap function and the generalized thickness function.

Finally, for sufficiently large nuclei, the overlap functions rapidly and effectively restricts the support of the integral to $\Omega_X \approx \Omega_X'$ such that
\begin{align}
\frac{\langle \Omega_A| \Omega_A'\rangle_{\rm GOA}}{|N_X|_{\rm GOA}^2} &\xrightarrow[\text{ large }A,B]{\text{asymptotically}} \frac{1}{4\pi} \delta^{(2)}(\Omega_X-\Omega_X')\,,\\
P_{AB}^{\rm inel}(\mathbf{b})  & \xrightarrow[\text{ large }A,B]{\text{asymptotically}} \int \frac{d\Omega_A}{4\pi} \int \frac{d\Omega_B}{4\pi}\nonumber\\
&\times \Big\{1- \left[1-P_{\Omega_A\Omega_A;\Omega_B\Omega_B}(\mathbf{b})\right]^{AB}\Big\}\,.
\label{eq:coll-classical}
\end{align}
Now it recovers conventional MC Glauber models with a classical-rotor picture, illustrated on the right of Fig.~\ref{fig:geo}.

{\bf \noindent The phase-space density approximation---}
If the microscopic overlap function and intrinsic-state wave function are available, the extended Glauber model can be applied directly. For a practical phenomenological implementation of Eq.~\eqref{eq:New-Glauber-2}, we make further phase-space density approximations. As shown in details in Sec.~III of the Supplemental Material, we first represent each intrinsic state by a Slater determinant and reduce the problem to computing overlaps of one-body density operators. Representing the density operators in phase space and performing a gradient expansion then allow the generalized thickness function and the overlap function to be expressed solely in terms of the intrinsic one-body phase-space distribution. Here, we summarize the final results.

Let $f_X(\mathbf r,\mathbf p;\Omega_X)$ denote the single-particle phase-space distribution in the intrinsic state at an orientation $\Omega_X$, and define the three-dimensional phase-space overlap as
\begin{align}
\mathcal F_X(\mathbf r, \mathbf p;\Omega_X,\Omega_X')  \equiv
\frac{\sqrt{f_X(\mathbf r,\mathbf p;\Omega_X)
f_X(\mathbf r,\mathbf p;\Omega_X')}}{(2\pi)^3}\,.
\end{align}
Then, the approximated generalized thickness function and overlap function are 
\begin{align}
&\hskip-.25em T_X(\bfx;\Omega_X,\Omega_X')
\approx
\frac{
\int dz\,d^3\mathbf p\, \mathcal F_X(\mathbf r, \mathbf p;\Omega_X,\Omega_X')
}{
\int d^3\mathbf r\,d^3\mathbf p\,\mathcal F_X(\mathbf r, \mathbf p;\Omega_X,\Omega_X')
}\,,
\label{eq:approx1}
\\
&\hskip-.25em \langle\Omega_X|\Omega_X'\rangle_{\rm GOA}
\approx
\left[
\frac{
\int d^3\mathbf r\,d^3\mathbf p\,\mathcal F_X(\mathbf r, \mathbf p;\Omega_X,\Omega_X')
}{
\int d^3\mathbf r\,d^3\mathbf p\,\mathcal F_X(\mathbf r, \mathbf p;\Omega_X,\Omega_X)
}
\right]^{\frac{X}{2}}\,.
\label{eq:approx2}
\end{align}
where $z$ is the longitudinal coordinate of $\mathbf r=(\bfx, z)$.

We model the phase-space distribution by a local Fermi-gas ansatz
\begin{align}
f_X(\mathbf r,\mathbf p; \hat{z})
=
1/\left(1+e^{
[\epsilon_p-\epsilon_{F,X}(\mathbf r)]/\Delta
}\right)\,.
\end{align}
$\Delta$ is a Fermi-surface smearing parameter.
The local Fermi energy $\epsilon_{F,x}(\mathbf r)$ is fixed by requiring the momentum-integrated distribution reproducing the Woods--Saxon density in the intrinsic frame,
\begin{align}
g\int\frac{d^3p}{(2\pi)^3}
f_X(\mathbf r,\mathbf p; \hat{z})
=
\frac{n_0}{
1+e^{
[r-R_X(\theta,\phi)]/a_X}
}\,.
\end{align}
Here $g$ accounts for spin and isospin degeneracies, and
\begin{align}
R_X(\theta,\phi)
=
R_{0,X}
\Big[
1+
\beta_{2,X}Y_{20}(\theta,\phi)+\cdots
\Big].
\end{align}
Therefore, the required inputs are reduced to the radius, deformation parameters, spatial diffuseness $a_X$, that parametrize the intrinsic state, and $\Delta$.

\begin{figure}[!t]
\centering
\includegraphics[scale=.68]{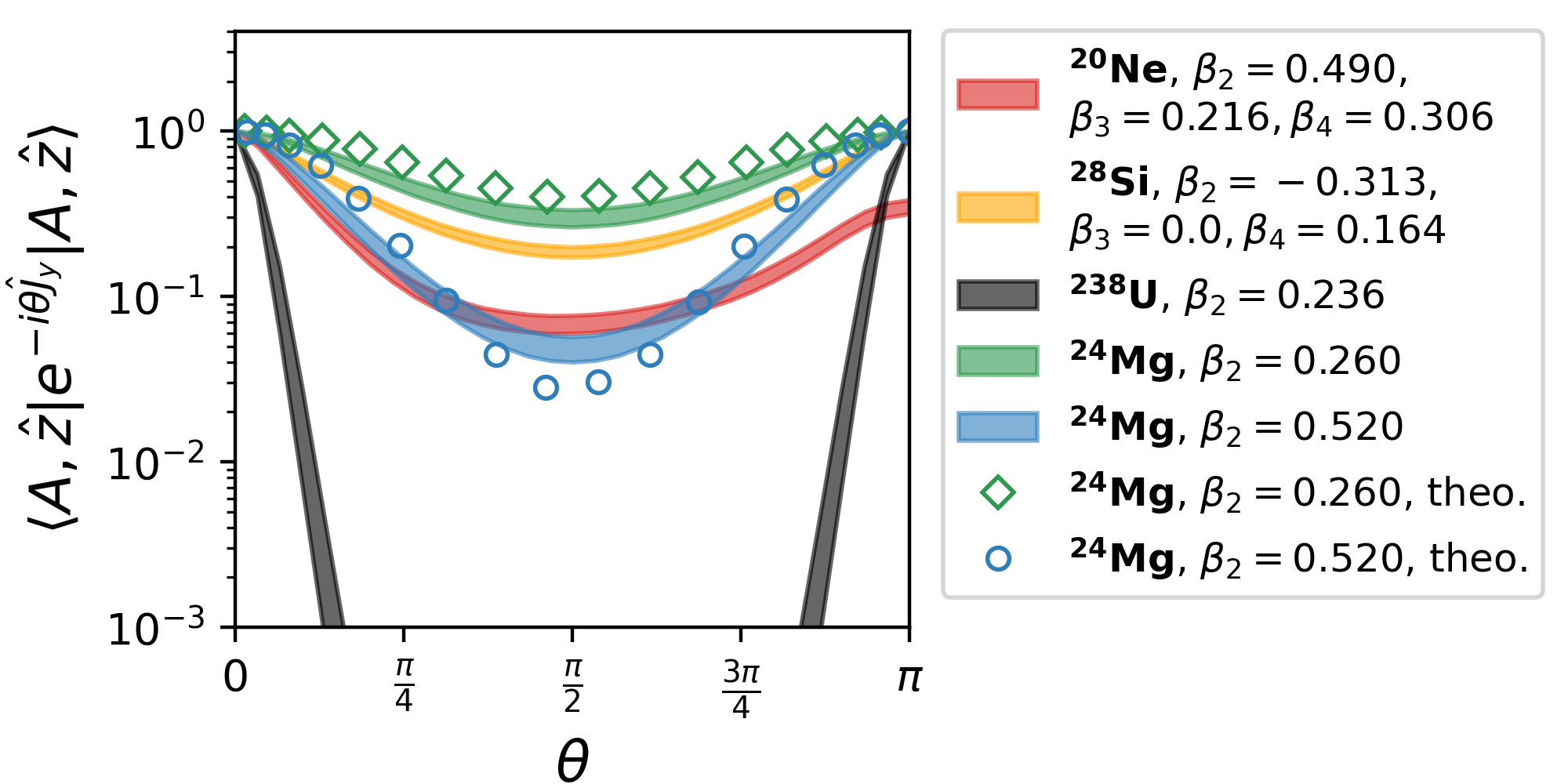}
\caption{Overlapping functions obtained from the phase-space overlap approximation in Eq.~\eqref{eq:approx2} for various deformed nuclei. The band widths reflect the variation of $\Delta$ from 0.25 to 0.5 MeV. For ${}^{24}\mathrm{Mg}$, we compare the approximation to the results obtained from the microscopic nuclear-structure calculation of Ref.~\cite{Yao:2009ry} for two different deformations (green diamond symbols for $\beta_2 = 0.26$ and blue circles for $\beta_2 = 0.52$).  }
\label{fig:Coverlap}
\end{figure}

In Fig.~\ref{fig:Coverlap}, using $\Delta=0.25$ to $0.5~\mathrm{MeV}$, we calculate the overlap functions for Ne, Mg, Si, and U with deformation parameters from Refs.~\cite{Moller:2015fba,Loizides:2025ule,Yao:2009ry}. The rapid decorrelation of ${}^{238}\mathrm U$ supports the classical-rotor approximation for heavy nuclei, whereas light nuclei retain substantial coherence over a broad angular range. In particular, for two deformations of ${}^{24}\mathrm{Mg}$, our simplified model agrees well with the microscopic calculations of Ref.~\cite{Yao:2009ry}, except near $\theta=\pi/2$, where the overlap is already small.

{\bf \noindent Phenomenology impact on QGP eccentricity.---} Using the main formula in Eq.~\eqref{eq:New-Glauber-2},~\eqref{eq:coll-pab},~\eqref{eq:TX} and the phase-space density approximations in Eqs.~\eqref{eq:approx1},~\eqref{eq:approx2}, we now directly sample the nucleon positions and estimate the energy deposition $\mathcal{E}(\bfx)$ to perform phenomenological calculations of the eccentricities of the produced QGP.

However, obtaining $\mathcal{E}(\bfx)$ from first principles based on the distribution of participant nucleons remains challenging. Here, we use the TRENTo proxy model~\cite{Moreland:2014oya}, which is sufficiently flexible to estimate the associated theoretical uncertainties and has been found to provide a good description of a wide range of high-energy nuclear collision data~\cite{JETSCAPE:2020mzn}.
Given the sampled participant nucleons, the TRENTo parametrization of energy deposition is
\begin{align}
\mathcal{E}(\bfx) &=
\mathcal{E}_0 \times \left( \frac{n_A^p(\bfx) + n_{B}^p(\bfx)}{2}\right)^{1/p},\\
n_A(\bfx) &= \sum_{i\in \textrm{participants}} \frac{\gamma_i}{2\pi w^2} \exp\Big(-\frac{(\bfx-\bfx_i)^2}{2w^2}\Big),
\end{align}
with $n_B(\bfx)$ defined analogously. $p$ controls the deposition ansatz and has been tightly constrained to $p\approx 0$~\cite{JETSCAPE:2020mzn}; $w$ controls the transverse smearing of the deposited energy around the participant-nucleon position $\bfx_i$; and $\gamma_i$ are independent random variables drawn from a unit-mean Gamma distribution with variance $1/k$, governing fluctuations in the energy deposition. Finally, $\mathcal{E}_0$ is a normalization that is irrelevant for computing eccentricities.

After the center-of-mass re-centering of $\mathcal{E}(\bfx)$, the eccentricities are computed as
\begin{align}
\epsilon_n = \frac{\int \mathcal{E}(\bfx) \, e^{in\phi} \, |\bfx|^n \, d^2\bfx}{\int \mathcal{E}(\bfx) \, |\bfx|^n \, d^2\bfx},
\end{align}
where $\phi$ is the azimuthal angle of $\bfx$.

To assess the impact of quantum effects, we compare eccentricities $\epsilon_n$ computed using participant nucleons sampled from Eq.~\eqref{eq:New-Glauber-2} (labeled ``Q'') to those obtained in the classical limit of Eq.~\eqref{eq:coll-classical} (labeled ``C''). 

\begin{figure}[!t]
\centering
\includegraphics[scale=.67]{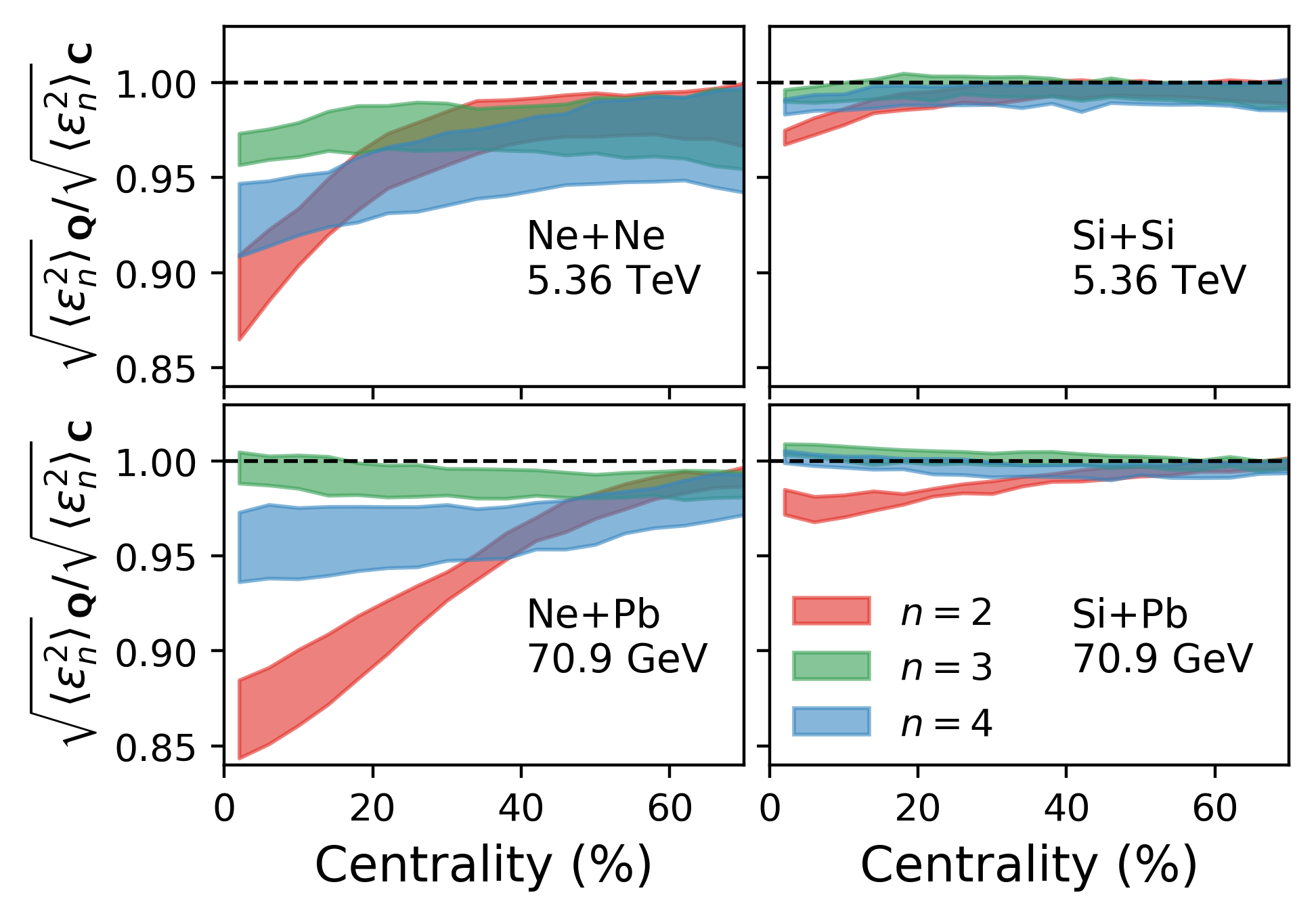}
\caption{Ratio of the root-mean-square eccentricities $\sqrt{\langle|\epsilon_n|^2\rangle}$ obtained from the quantum-rotor and classical-rotor treatments as a function of centrality. The bands include uncertainties from variations of both the TRENTo energy-deposition parameters and the approximate overlap function through the parameter $\Delta$.}
\label{fig:ecc}
\end{figure}

In Fig.~\ref{fig:ecc}, we show the quantum-to-classical ratio $\sqrt{\langle|\epsilon_n|^2\rangle}_Q/ \sqrt{\langle|\epsilon_n|^2\rangle}_C$ for $n=2,3,4$ as a function of centrality. The bottom row corresponds to collisions of light nuclei with ${}^{208}$Pb, while the top row shows collisions of identical light nuclei. The bands represent uncertainties in the energy-deposition model, estimated by varying the TRENTo parameters over $p\in[-1,1]$, $w\in[0.6,0.8]$ fm, and $k\in[1/2,2]$, together with the uncertainty in the approximate overlap function obtained by varying $\Delta\in[0.25,0.5]$ MeV. These combined uncertainties affect the quantum-to-classical correction by less than $4\%$.

For the strongly prolate ${}^{20}$Ne, with the same deformation parameters as in Fig.~\ref{fig:Coverlap}, quantum treatment reduces $\sqrt{\langle|\epsilon_2|^2\rangle}$ by $13\%$ in the 0-10\% central Ne+Pb collisions relative to the classical-rotor result. This reduction arises because each generalized thickness function depends on two orientations, thereby smearing the imprint of the intrinsic deformation on the collision geometry. For the oblate deformed $^{28}$Si nucleus, the reduction is less than 3\%. 

The impact on $\epsilon_2$ is less pronounced in central symmetric collisions (Ne+Ne and Si+Si) than in asymmetric light-heavy collisions. In the former case, a large number of participants requires both nuclei to occupy comparable configurations, which weakens the decorrelation effect. 

Therefore, the quantum versus classical-rotor modeling could influence the interpretation of (Ne+Ne)/(O+O) flow ratio measurements~\cite{ALICE:2025luc,ATLAS:2025nnt,CMS:2025tga}, which require percent-level precision in comparisons with theory~\cite{Giacalone:2024ixe}. Recent studies using microscopic nuclear structure inputs or analytic models~\cite{Bofos:2026nmg,Xu:2026llz} have also reported similar magnitude of quantum corrections on eccentricity.

{\noindent \bf Summary.---}We extend the conventional classical-rotor implementation of the MC Glauber model by introducing a modified nucleon-position sampling procedure that incorporates the constraints imposed by the rotational symmetry of the ground states of deformed even-even nuclei. Under the independent-pair approximation and the GOA, this procedure requires information absent from the conventional model: an overlap function between intrinsic states at different orientations and a generalized thickness function that depends on off-diagonal elements in the basis of orientations. Within this framework, the predicted QGP geometry differs from that obtained with the classical-rotor treatment. The latter is recovered when the overlap between distinct orientations becomes sufficiently localized, e.g., in the macroscopic limit.

For a practical implementation, we employ a simplified phase-space density approximation to estimate the overlap function and the generalized thickness function. Relative to the classical-rotor treatment, our approximate quantum-rotor treatment reduces $\sqrt{\langle|\epsilon_2|^2\rangle}$ by $13\%$ in the $0$--$10\%$ most central ${}^{20}\mathrm{Ne}+{}^{208}\mathrm{Pb}$ collisions and by approximately $10\%$ in central ${}^{20}\mathrm{Ne}+{}^{20}\mathrm{Ne}$ collisions. These effects could influence the quantitative interpretation of recent light-ion collision data. In the future, the framework can be extended to additional generalized coordinates such as $\{\Omega,\beta_2,\beta_3,\ldots\}$, which are relevant to studies of nuclear shape fluctuations and coexistence in HICs~\cite{Zhao:2024lpc,Xu:2025cgx,Henderson:2025scq}.

{\bf\noindent Acknowledgments.---} The author would like to thank Giuliano Giacalone, Jiangyong Jia, Huichao Song, Qun Wang, Hao-jie Xu, Shuai Yang, Jiangming Yao, Chunjian Zhang, Pengwei Zhao, Weibin Zhao, Bo Zhou, and for discussions and comments on the manuscript. Special thanks to Jiangyong Jia, Huichao Song, Chunjian Zhang for their kind invitation to a series of workshops that brought together the nuclear structure and relativistic heavy-ion collision communities, fostering the development of this work. The authors acknowledge the use of an AI-assisted language tool to improve the clarity and grammar of the manuscript. This work is supported by NSFC under Grant No. 12575140.

\onecolumngrid

\appendix

\section{On the difference between SD, NSD, ND cross sections}
Here we comment on the subtle distinction between the minimum-bias (MB) event samples measured experimentally and the non-diffractive (ND) events calculated in Eq.~(3) of the main Letter.

In ultrarelativistic nuclear collisions, the minimum-bias (MB) trigger is designed to record essentially all hadronic collisions that produce observable final-state particles~\cite{ALICE:2013hur}. We identify this sample with non-single-diffractive (NSD) events, namely events excluding single diffraction. The difference between these two event selections matters only in very peripheral collisions.

Experimentally, the diffractive events relevant here fall into two categories:
\begin{itemize}
    \item Single-diffractive (SD) events, in which one incoming hadron or nucleus diffractively dissociates, while the other remains intact or is only weakly excited.
    \item Double-diffractive (DD) events, in which both incoming hadrons or nuclei dissociate, with the particles produced by the two diffractive systems separated by a rapidity gap according to an experiment-specific criterion.
\end{itemize}
The difference between ND and NSD is whether DD events are included, $\sigma_{\rm NSD} = \sigma_{\rm ND}+\sigma_{\rm DD},
$, where the much smaller central-diffractive contribution has been neglected. This distinction between ND and NSD is relevant mainly for very peripheral, proton--proton-like collisions involving only one or a few binary interactions. For collisions with appreciable nuclear overlap, particularly the central collisions of interest for studying deformation effects, additional binary interactions always produce particles across rapidity and fill the gap required to identify a DD topology. The DD contribution to the selected nuclear event sample then becomes negligible, making the difference between ND and NSD irrelevant for the collision geometries considered here. We therefore use the ND cross
section throughout this work.

\section{The independent-pair approximation of the many-body collision term}

Consider two nuclei $A$ and $B$, whose centers are separated by the transverse impact parameter $\mathbf b$. Let $\mathbf{x}_a$ and $\mathbf{x}_b$ denote the transverse coordinates of nucleons $a\in A$ and $b\in B$, respectively, relative to the centers of their nuclei. For compactness, we denote the normalized transition matrix element of an operator $\hat O$ by
\begin{align}
\overline{O} \equiv
\frac{ \langle\Omega_A,\Omega_B|\hat O|\Omega_A',\Omega_B'\rangle}{\langle\Omega_A|\Omega_A'\rangle \langle\Omega_B|\Omega_B'\rangle}.
\label{eq:transition-expectation}
\end{align}
For symmetry-restored nuclear states, the inelastic collision probability at fixed impact parameter is
\begin{align}
P_{AB}^{\rm inel}(\mathbf b) ={}& \int d\Omega_A\,d\Omega_A' \int d\Omega_B\,d\Omega_B'\,
\frac{\langle\Omega_A|\Omega_A'\rangle}{|N_A|^2}
\frac{\langle\Omega_B|\Omega_B'\rangle}{|N_B|^2}
\left[1-\mathcal G_{AB}\left(\mathbf b;\Omega_A,\Omega_A';\Omega_B,\Omega_B'\right)\right],
\label{eq:IAB-definition}
\end{align}
where $N_X$ is the normalization of the corresponding symmetry-restored state, and
\begin{align}
\mathcal G_{AB}\left(\mathbf b;\Omega_A,\Omega_A';\Omega_B,\Omega_B'\right)\equiv\overline{\prod_{a=1}^{A}\prod_{b=1}^{B}\left[1-C\left(\hat{\mathbf x}_a-\hat{\mathbf x}_b-\mathbf b\right)\right]}.
\label{eq:GAB-normalized}
\end{align}
For the time-reversal-invariant even--even intrinsic states considered in this Letter, $\mathcal G_{AB}$ is real. 

We analyze $\mathcal G_{AB}$ using a cumulant expansion. The expansion contains binary collisions between the same pair $(a,b)$, and mixed contributions involving distinct pairs $(a,b)\neq (a',b')$. We will resum the former exactly and neglect the latter, thereby discarding multi-nucleon correlations in the intrinsic state that couple different binary collisions. Correlations generated by the coherent superposition of intrinsic states at different orientations are nevertheless retained through the integrations over $(\Omega_X,\Omega_X')$ and the corresponding generalized thickness functions.

To separate correlations between distinct binary pairs, let $i=(a,b)$ label the $N=AB$ binary pairs and define
\begin{align}
y_i\equiv\ln(1-C_i).
\end{align}
By the multilinearity of cumulants, $\kappa_n(Y=\sum_i y_i)$ can be written as a sum of joint cumulants $\kappa(y_{i_1},\ldots,y_{i_n})$. When all pair labels are identical, this reduces to the ordinary cumulant $\kappa_n(y_i)\equiv\kappa(y_i,\ldots,y_i)$. The expansion therefore separates exactly into same-pair and mixed contributions:
\begin{align}
\mathcal G_{AB}\left(\mathbf b;\Omega_A,\Omega_A';\Omega_B,\Omega_B'\right)&= \overline{\prod_{i=1}^{AB}(1-C_i)}\nonumber\\
&=\exp\left\{\sum_{n=1}^{\infty}\frac{1}{n!}\sum_{i_1,\ldots,i_n}\kappa(y_{i_1},\ldots,y_{i_n})\right\}\nonumber\\
&=\exp\left\{\sum_{i=1}^{AB}\sum_{n=1}^{\infty}\frac{\kappa_n(y_i)}{n!}+W_{\rm mix}\right\},
\end{align}
where
\begin{align}
W_{\rm mix}\equiv\sum_{n=2}^{\infty}\frac{1}{n!}\sum_{i_1\cdots i_n}\hskip-.3em{}^{\prime}\,\kappa(y_{i_1},\ldots,y_{i_n}),
\end{align}
and the prime denotes terms containing at least two distinct pair labels. For each fixed pair, the same-pair cumulants resum exactly:
\begin{align}
\sum_{n=1}^{\infty}\frac{\kappa_n(y_i)}{n!} = \ln\overline{e^{y_i}} = \ln\left(1-\overline{C_i}\right).
\end{align}

The independent-pair approximation neglects only the mixed connected cumulants, $W_{\rm mix}\approx0$, while retaining all same-pair contributions. Since we do not distinguish protons and neutrons, $\overline{C_{ab}}\equiv P_{\Omega_A\Omega_A';\Omega_B\Omega_B'}(\bfb)$ is the same for every pair. We therefore obtain
\begin{align}
\mathcal G_{AB} \left(\mathbf b;\Omega_A,\Omega_A';\Omega_B,\Omega_B'\right)\approx\left[1-P_{\Omega_A\Omega_A';\Omega_B\Omega_B'}(\bfb)\right]^{AB}.
\end{align}

Now, we can write down the definition of the generalized thickness function. Since each pair of collision is the same, we can use the field operator representation of the nucleon creation and annihilation operators, and average the indices over all pairs.
\begin{align}
\frac{1}{AB}\sum_{a=1}^{A}\sum_{b=1}^{B}\hat C_{ab}(\mathbf b)=\frac{1}{AB}\int d^3r_A\,d^3r_B\,C\left(\mathbf x_A-\mathbf x_B-\mathbf b\right)\hat\psi_A^\dagger(\mathbf r_A)\hat\psi_B^\dagger(\mathbf r_B)\hat\psi_B(\mathbf r_B)\hat\psi_A(\mathbf r_A),
\label{eq:C-second-quantized}
\end{align}
where $\mathbf r_X=(\mathbf x_X,z_X)$. Because the incoming state factorizes between nuclei $A$ and $B$, the collision probability between two nucleons factorizes into the binary collision profile and the generalized thickness function of each nuclesu
\begin{align}
P_{\Omega_A\Omega_A';\Omega_B\Omega_B'}(\bfb) = &\int d^2x_A\,d^2x_B\, C\left( \mathbf x_A-\mathbf x_B-\mathbf b\right)T_A\left(\mathbf x_A;\Omega_A,\Omega_A'\right)T_B\left(\mathbf x_B;\Omega_B,\Omega_B'\right)
\label{eq:K1-thickness}
\end{align}
where the generalized thickness function is defined by
\begin{align}
T_X\left(\mathbf x;\Omega_X,\Omega_X'\right)\equiv\frac{1}{X}\int dz\,\frac{\langle\Omega_X|\hat n_X(\mathbf x,z)|\Omega_X'\rangle}{\langle\Omega_X|\Omega_X'\rangle}\,, \\
\hat{n}(\mathbf x,z) = \hat\psi_X^\dagger(\mathbf x,z)\hat\psi_X(\mathbf x,z)\,,\qquad X=A,B.
\label{eq:generalized-thickness-definition}
\end{align}
It is also normalized $\int d^2\bfx\,
T_X\left(\mathbf x;\Omega_X,\Omega_X'\right)= 1$.

On the diagonal, $\Omega_X=\Omega_X'$, $T_X$ reduces to the conventional nuclear thickness divided by $X$. For different orientations, $T_X$ is an off-diagonal transition thickness rather than a probability density and may in general be complex. For the even--even intrinsic Slater determinants considered here, complete Kramers pairs are occupied and the intrinsic state is invariant under time reversal. Because spatial rotations commute with time reversal and the local scalar density is time-reversal even, both the overlap and the local transition-density matrix element are real. Consequently $T_X(\mathbf x;\Omega_X,\Omega_X')\in\mathbb R$. It is nevertheless not necessarily positive, except under the additional conditions discussed below.

\section{Approximate calculations of $T_X(\bfx; \Omega_X, \Omega_X')$ and $\langle \Omega_X|\Omega_X'\rangle_{\rm GOA}$}
\label{sec:practical-reduced-thickness}

The generalized thickness defined in Eq.~\eqref{eq:generalized-thickness-definition} requires the off-diagonal many-body transition density. We simplify this input in two stages. First, since the independent-pair approximation adopted above already neglects multi-nucleon correlations in the intrinsic state that, we consistently represent each intrinsic state by a single Slater determinant. We then express the exact transition density in terms of the projectors onto the two occupied single-particle subspaces and derive the one-body density-overlap approximation. Second, we replace the remaining nonlocal density matrices by smooth phase-space distributions within a local Fermi-gas model. 

\subsection{The Slater determinant approximation of the intrinsic state}

We assume that the intrinsic state with orientation $\Omega_X$ of nucleus $X=A,B$ is represented by a Slater determinant $\langle \{\mathbf r_m\}|\Omega_X\rangle = \det\{\phi_n(\mathbf r_m)\}$, where $\phi_n(\mathbf{r})=\langle \mathbf r|n\rangle$ are the single-particle orbits. Its one-body density matrix is 
\begin{align}
\hat{\rho}_X = \sum_{n=1}^X \frac{1}{X} |n\rangle\langle n|
\end{align}
for $X=A, B$. $n$ sums over all the occupied single-particle states. The projector onto subspace of the occupied states are related to the one-body density matrix by 
\begin{align}
\hat P_X &= X\hat\rho_X.
\end{align}
For another orientation $\Omega'$ that is related to orientation $\Omega$ by a rotation $\hat U_X=\exp\{-i \theta \vec{n}\cdot \hat{J}_X\}$, the state, projector and density matrix in the $\Omega'$ orientated configuration are related by
\begin{align}
|n'\rangle &= U_X|n\rangle\,,\\
\hat P_X'&=\hat U_X\hat P_X\hat U_X^\dagger\,, \\
\hat\rho_X'&=\hat U_X\hat\rho_X\hat U_X^\dagger\,.
\end{align}

Now, we use these projectors or one-body density operators to express the generalized thickness function. Let
\begin{align}
(S_X)_{mn}\equiv\langle m|n'\rangle,
\end{align}
be the overlap matrix of the occupied orbitals from the two intrinsic states at two different orientations $\Omega$ and $\Omega'$. Then, the overlap between the two configuration can be expressed as $\langle\Omega_X|\Omega_X'\rangle=\det S_X$, and $(S_X S^\dagger_X)_{mn} = \langle m|\hat P_X\hat P_X'\hat P_X |n\rangle$. Then, the inverse of $S_X$ can be written as
\begin{align}
(S_X^{-1})_{mn} = (S_X^\dagger)_{mk}((S_XS_X^\dagger)^{-1})_{kn} = (S_X^\dagger)_{mk} \langle k|(\hat P_X\hat P_X'\hat P_X)^{-1}|n\rangle \,.
\end{align}

With these preparation, the transition-type of expectation of the one-body density operator in Eq.~\ref{eq:generalized-thickness-definition} can be expressed as
\begin{align}
\frac{\langle\Omega_X|
\hat\psi_X^\dagger(\mathbf r)
\hat\psi_X(\mathbf r)
|\Omega_X'\rangle}
{\langle\Omega_X|\Omega_X'\rangle}
&=
\frac{\sum_{nm} \phi_n^*(\mathbf r)\phi_m'(\mathbf r)\operatorname{Cof}(S_X)_{nm}}{\det S_X}
=
\sum_{mn} \langle \mathbf r|m'\rangle   (S_X^{-1})_{mn} \langle n|\mathbf r\rangle \nonumber\\
&= \sum_{mnk}\langle \mathbf r|n'\rangle  (S_X^\dagger)_{nk} \langle k|(\hat P_X\hat P_X'\hat P_X)^{-1}|m \rangle \langle m |\mathbf r\rangle  \nonumber\\
&= \sum_{mnk}\langle \mathbf r|m'\rangle  \langle m'|k\rangle \langle k|(\hat P_X\hat P_X'\hat P_X)^{-1}|n \rangle \langle n |\mathbf r\rangle  \nonumber\\
&=  \langle \mathbf r| \hat{P}_X' \hat{P}_X (\hat P_X\hat P_X'\hat P_X)^{-1} \hat{P}_X |\mathbf r\rangle  
\end{align}

Therefore the generalized transition
thickness, can then be written entirely in terms of
$\hat P_X$ and $\hat P_X'$, without choosing individual occupied
orbitals:
\begin{align}
T_X\left(\mathbf x;\Omega_X,\Omega_X'\right)&=\frac{1}{X}\int dz\,\left\langle\mathbf x,z\left|\hat P_X'\hat P_X(\hat{G}_X)^{-1}\hat P_X\right|\mathbf x,z\right\rangle,\\
\hat{G}_X &\equiv \hat P_X\hat P_X'\hat P_X
\label{eq:exact-thickness-projectors}
\end{align}
So $\hat{G}_X$ is the operator corresponding to the overlapping matrix product $(S_X S^\dagger_X)_{mn}$ in $|n\rangle$ basis .

\subsection{One-body density operator overlap approximation}
The next step of approximation concerns the overlapping matrix $(S_X S^\dagger_X)_{mn}$, or equivalently speaking, the operator $\hat{G}_X$.
\begin{align}
\hat G_X=\hat P_X\hat P_X'\hat P_X=\hat P_X\hat U_X\hat P_X\hat U_X^\dagger\hat P_X\,.
\end{align}
It measures the violation of unitary when the rotation $\hat U_X$ is restricted to the occupied subspace. Indeed,
\begin{align}
\hat G_X=\hat P_X-\hat P_X\hat U_X(1-\hat P_X)\hat U_X^\dagger\hat P_X,
\end{align}
where the second term describes the leakage of rotated occupied states into the unoccupied subspace. We retain the average leakage but neglect
its dependence on the direction within the occupied subspace, so that
\begin{align}
\hat G_X \approx \frac{\operatorname{Tr}\hat G_X}{X}\hat P_X = \frac{\operatorname{Tr}\left(\hat P_X'\hat P_X\right)}{X}\hat P_X
\end{align}
Consequently, with the inverse understood within the occupied subspace,
\begin{align}
\hat G_X^{-1} \approx \frac{X}{ \operatorname{Tr} \left(\hat P_X'\hat P_X\right)}\hat P_X.
\end{align}

Substitution into Eq.~\eqref{eq:exact-thickness-projectors} gives
\begin{align}
T_X\left(\mathbf x;\Omega_X,\Omega_X'\right)&\approx\frac{\displaystyle\int dz\,\langle\mathbf x,z|\hat P_X'\hat P_X|\mathbf x,z\rangle}{\displaystyle\operatorname{Tr}\left(\hat P_X'\hat P_X\right)}.
\label{eq:TX-projector-overlap}
\end{align}

Further using the relation
$\sqrt{\hat\rho_X}=\frac{\hat P_X}{\sqrt X},\sqrt{\hat\rho_X'}=\frac{\hat P_X'}{\sqrt X}$,
it can also be expressed approximately in terms of the one-body density matrices, and we arrive at the one-body density operator overlap approximation of the generalized thickness function
\begin{align}
T_X \left(\bfx;\Omega_X,\Omega_X'\right)=\frac{\displaystyle\int dz\,\left\langle\bfx,z\left|\sqrt{\hat\rho_X'}\sqrt{\hat\rho_X}\right|\bfx,z\right\rangle}{\displaystyle\operatorname{Tr}\left(\sqrt{\hat\rho_X'}\sqrt{\hat\rho_X}\right)}.
\label{eq:TX-density-overlap}
\end{align}
This is the first practical reduction: it eliminates the occupied-space inverse but still requires the complete one-body density matrices. The next subsection removes this remaining microscopic input.

\subsection{The phase-space density representation and the local Fermi-gas model}

We now perform the Wigner transformation and convert the one-body density matrix to its phase-space representation, and the latter will be approximated by a local Fermi-gas model to be used in phenomenology. The Wigner transformation of a single-particle operator $\hat A$ 
\begin{align}
\mathcal{W}[\hat{A}]\equiv\int d^3\boldsymbol{\delta}\,e^{-i\mathbf p\cdot\boldsymbol{\delta}}\left\langle\mathbf r+\frac{\boldsymbol{\delta}}{2}\left|\hat A\right|\mathbf r-\frac{\boldsymbol{\delta}}{2}\right\rangle\,.
\label{eq:Wigner-definition}
\end{align}
Define 
\begin{align}
s_X(\mathbf r,\mathbf p) \equiv \mathcal W\left[\sqrt{\hat\rho_X}\right]\,,\quad
s_X'(\mathbf r,\mathbf p)\equiv\mathcal W\left[\sqrt{\hat\rho_X'}\right]\,,
\end{align}
The Wigner transform of the one-body density operator overlap is the Moyal product, which has a gradient expansion
\begin{align}
\mathcal W\left[\sqrt{\hat\rho_X'}\sqrt{\hat\rho_X}\right]=s_X's_X+\frac{i}{2}\left\{s_X',s_X\right\}_{\rm PB}+\mathcal O(\partial^2 s_X \partial^2 s_X').
\label{eq:Moyal-expansion}
\end{align}
where the classical Poisson bracket is
\begin{align}
\{f,g\}_{\rm PB}\equiv\nabla_{\mathbf r}f\cdot\nabla_{\mathbf p}g-\nabla_{\mathbf p}f\cdot\nabla_{\mathbf r}g.
\end{align}
It follows that
\begin{align}
\left\langle\mathbf r\left|\sqrt{\hat\rho_X'}\sqrt{\hat\rho_X}\right|\mathbf r\right\rangle=g_X\int\frac{d^dp}{(2\pi)^d}\left[s_X'(\mathbf r,\mathbf p)s_X(\mathbf r,\mathbf p)+\frac{i}{2}\left\{s_X',s_X\right\}_{\rm PB}+\mathcal O(\partial^2 s_X \partial^2 s_X')\right],
\label{eq:local-density-Moyal}
\end{align}
where $g_X$ is the internal spin-isospin degeneracy. Here we have use that for a time-reversal-invariant even--even nucleus, the phase-space distributions are even under $\mathbf p\rightarrow-\mathbf p$. The Poisson-bracket term in Eq.~\eqref{eq:local-density-Moyal} is then odd in $\mathbf p$ and vanishes after momentum integration. The first non-vanishing correction to the local density therefore appears at second order in the gradient expansion.

The final step is use a Fermi-gas model for the phase-space representation of the one-body density matrix
\begin{align}
\mathcal W[\hat\rho_X]
\approx
\frac{g}{X}
f_X(\mathbf r,\mathbf p;\Omega_X).
\end{align}
which is normalized to
\begin{align}
g\int d^3x\int\frac{d^3p}{(2\pi)^3}f_X(\mathbf r,\mathbf p;\Omega_X)=X.
\label{eq:FX-normalization}
\end{align}
where $g$ is the degeneracy factor accounting for spin and isospin.

For a smoothly varying phase-space distribution, the leading term in gradient
 relates
\begin{align}
s_X(\mathbf r,\mathbf p)=\mathcal W[\sqrt{\hat\rho_X}]\approx\sqrt{\frac{g}{X}f_X(\mathbf r,\mathbf p;\Omega_X)}.
\label{eq:sqrt-rho-Wigner}
\end{align}
Substitution into Eq.~\eqref{eq:local-density-Moyal} gives
\begin{align}
\left\langle\mathbf r\left|\sqrt{\hat\rho_X'}\sqrt{\hat\rho_X}\right|\mathbf r\right\rangle\approx\frac{g}{X}\int\frac{d^3\mathbf{p}}{(2\pi)^3}\sqrt{f_X(\mathbf r,\mathbf p;\Omega_X)f_X(\mathbf r,\mathbf p;\Omega_X')}.
\label{eq:local-phase-space-overlap}
\end{align}

Finally, we use a local Fermi-gas model with occupation number
\begin{align}
f_X(\mathbf x,\mathbf p; \Omega_X=\hat{z}) = \frac{1}{1+\exp\left[\frac{\mathbf p^2-p_{F,X}^2(\mathbf x)}{2M_N\Delta}\right]},
\label{eq:local-Fermi-distribution}
\end{align}
where $\Delta$ controls the smearing of the local Fermi surface. In the numerical implementation, we take $\Delta=0.25$ to $0.5$~MeV, which are found to provide a reasonable agreement with microscopic calculations~\cite{Yao:2009ry}. The local Fermi momentum is determined from the intrinsic density by
\begin{align}
n_X(\mathbf x)=g_X\int\frac{d^3p}{(2\pi)^3}f_X^{(0)}(\mathbf x,\mathbf p)\,,
\label{eq:pF-implicit}
\end{align}
while the intrinsic spatial density is parametrized by a deformed Woods--Saxon distribution,
\begin{align}
n_X(\mathbf x)=
\frac{n_{0,X}}{1+\exp\left[\frac{r-R_X(\vartheta,\varphi)}{a_X}
\right]}\,,\quad R_X(\theta,\phi)=R_{0,X}\left[1+\sum_{\ell,m}\beta_{\ell m}^{X}Y_{\ell m}(\theta,\phi)\right]\,,
\label{eq:deformed-radius}
\end{align}
and the normalization $n_{0,X}$ is fixed by
$\int d^3x\,n_X(\mathbf x)=X$.

Therefore, this approximation is convenient in phenomenological application: once the intrinsic Woods--Saxon density, including its radius, surface diffuseness, and deformation parameters, is specified, Eq.~\eqref{eq:TX-density-overlap} and Eq.~\eqref{eq:local-phase-space-overlap} determines the generalized thickness function required by the extended Monte Carlo Glauber model. 

\subsection{Final approximate expression for $T_X(\bfx; \Omega_X, \Omega_X')$ and $\langle \Omega_X|\Omega_X'\rangle_{\rm GOA}$}

Combining Eq.~\eqref{eq:TX-density-overlap} and Eq.~\eqref{eq:local-phase-space-overlap} and separate the dependence in transverse and longitudinal (beam) coordinates $\mathbf r=(\bfx, z)$, the generalized thickness function becomes
\begin{align}
T_X\left(\mathbf x;\Omega_X,\Omega_X'\right)=\frac{\displaystyle\int dz \int\frac{d^3\mathbf{p}}{(2\pi)^3}\sqrt{f_X(\mathbf x, z,\mathbf p;\Omega_X)f_X(\mathbf x, z,\mathbf p;\Omega_X')}}{\displaystyle\int dz \int d^2\bfx \int\frac{d^3\mathbf{p}}{(2\pi)^3}\sqrt{f_X(\mathbf x, z,\mathbf p;\Omega_X)f_X(\mathbf x, z,\mathbf p;\Omega_X')}}.
\label{eq:generalized-density-phase-space}
\end{align}
On the diagonal, this expression again reduces to the traditional treatment that is based on classical rotor model. 

Another input to the extended Monte Carlo Glauber model is the angular decorrelation function. We now show that the $X/2$-th power of the same one-body density overlap reproduces the many-body overlap to the accuracy of the Gaussian-overlap approximation (GOA), where $X=A, B$

For the time-reversal-invariant intrinsic state considered here, $\langle\hat J_y\rangle=0$, and the small-angle GOA gives
\begin{align}
\langle \Omega_X|\Omega_X'\rangle = \langle\Omega_X|e^{-i\theta\hat J_y}|\Omega_X\rangle\approx\exp\left[-\frac{\theta^2}{2}\langle\hat J_y^2\rangle\right].
\label{eq:GOA-overlap}
\end{align}
For a Slater determinant,
\begin{align}
\langle\hat J_y^2\rangle=\operatorname{Tr}\left(\hat P_X\hat j_y^2-\hat P_X\hat j_y\hat P_X\hat j_y\right),
\label{eq:Jy2-projector}
\end{align}
where $\hat P_X$ projects onto the occupied single-particle subspace. 

Another path to get this expression is to use the one-body density operator of this slater determinant
\begin{align}
\hat\rho_X=\frac{1}{X}\hat P_X,\qquad\hat\rho_X'=e^{-i\theta\hat j_y}\hat\rho_Xe^{i\theta\hat j_y},
\end{align}
and compute the following overlap
\begin{align}
d_1(\theta)\equiv\operatorname{Tr}\left(\sqrt{\hat\rho_X'}\sqrt{\hat\rho_X}\right).
\label{eq:d1-definition}
\end{align}
Expanding this expression to second order in $\theta$ and using
$\sqrt{\hat\rho_X}=\hat P_X/\sqrt X$, we obtain
\begin{align}
d_1(\theta)&=1-\frac{\theta^2}{X}\operatorname{Tr}\left(\hat P_X\hat j_y^2-\hat P_X\hat j_y\hat P_X\hat j_y\right)+\mathcal O(\theta^4) \nonumber\\
&=1-\frac{\theta^2}{X}\langle\hat J_y^2\rangle+\mathcal O(\theta^4).
\end{align}
Consequently,
\begin{align}
\left[d_1(\theta)\right]^{X/2}=\exp\left[\frac{X}{2}\ln d_1(\theta)\right]\approx\exp\left[-\frac{\theta^2}{2}\langle\hat J_y^2\rangle\right]
\label{eq:d1-GOA}
\end{align}
Thus, the same one-body density overlap determines both the generalized thickness function and the overlapping function. 

Using the phase-space representation and the Fermi-Gas model, the overlapping function is thus approximately calculated as
\begin{align}
\langle \Omega_X|\Omega_X'\rangle_{\rm GOA} 
&\approx\exp\left[-\frac{\theta^2}{2}\langle\hat J_y^2\rangle\right]\nonumber\\
&\approx \left[\frac{\displaystyle\int dz \int d^2\bfx \int\frac{d^3\mathbf{p}}{(2\pi)^3}\sqrt{f_X(\mathbf x, z,\mathbf p;\Omega_X)f_X(\mathbf x, z,\mathbf p;\Omega_X')}}{\displaystyle\int dz \int d^2\bfx \int\frac{d^3\mathbf{p}}{(2\pi)^3}f_X(\mathbf x, z,\mathbf p;\Omega_X)}\right]^{X/2}
\end{align}
Its effectiveness has been checked against microscopic calculations in Fig.~2 of the letter. The normalized de-correlation function under GOA is
\begin{align}
D_X(\Omega_X, \Omega_X')=\frac{\langle \Omega_X|\Omega_X'\rangle_{\rm GOA}}{|N_X|^2_{\rm GOA}}
\end{align}
with 
\begin{align}
|N_X|^2_{\rm GOA} = \int d\Omega_X \int d\Omega_X' \langle \Omega_X|\Omega_X'\rangle_{\rm GOA}.
\end{align}

\section{Decoherence in the transverse-position basis}
We provide a physical motivation for the probabilistic sampling of transverse nucleon positions in the Monte Carlo Glauber model.

Let $\mathbf R$ collectively denote the transverse positions of all nucleons in the two incoming nuclei. In the eikonal limit, these positions are effectively frozen during the collision, and the collision
operator is diagonal in the transverse-position basis~\cite{Miller:2007ri},
schematically,
\begin{align}
\langle\mathbf R'|\hat{\mathcal C}|\mathbf R\rangle = \delta(\mathbf R'-\mathbf R)\, \mathcal C(\mathbf R).
\end{align}
For an initial state
\begin{align}
|\Psi_{\rm in}\rangle = \int d\mathbf R\, \Psi_{\rm in}(\mathbf R)|\mathbf R\rangle,
\end{align}
the collision produces the final state
\begin{align}
|\Psi_{\rm out}\rangle
=
\int d\mathbf R\,
\Psi_{\rm in}(\mathbf R)|E_{\mathbf R}\rangle,
\end{align}
where $|E_{\mathbf R}\rangle$ denotes the many-particle final state produced from the incoming configuration $\mathbf R$. For inelastic collisions, the incoming nucleon degrees of freedom no longer remain identifiable in $|E_{\mathbf R}\rangle$.

For an operator $\hat O$ cooresponding to some coarse-grained quantity of the system after the collision (such as the energy momentum tensor required by hydrodynamic model, phase-space density required by kinetic theory, etc), its expectation value is
\begin{align}
\langle\hat O\rangle = \int d\mathbf R\,d\mathbf R'\, \Psi_{\rm in}(\mathbf R)\Psi^*_{\rm in}(\mathbf R')\, \langle E_{\mathbf R'}|\hat O|E_{\mathbf R}\rangle\,,
\end{align}
Because a relativistic inelastic collision produces many unobserved
degrees of freedom, interference between macroscopically distinguishable
final states is strongly suppressed. For the coarse-grained observables
relevant to the QGP geometry, we therefore assume
\begin{align}
\langle E_{\mathbf R'}|\hat O|E_{\mathbf R}\rangle
\approx
\delta(\mathbf R'-\mathbf R)\,
O(\mathbf R).
\end{align}
It then follows that
\begin{align}
\langle\hat O\rangle
\approx
\int d\mathbf R\,
|\Psi_{\rm in}(\mathbf R)|^2\,
O(\mathbf R).
\end{align}
Thus, although the incoming nucleon degrees of freedom no longer remain
in the final state, coarse-grained final-state observables depend only on
the diagonal transverse-position distribution
$\rho_{\rm in}(\mathbf R,\mathbf R)$. This provides the physical
motivation for sampling definite nucleon positions in the Monte Carlo
Glauber model.

Importantly, diagonality in the transverse-position basis does not imply diagonality in the generalized-coordinate basis. For the
symmetry-restored state
\begin{align}
|X\rangle = \frac{1}{N_X}\int d\Omega_X\,|\Omega_X\rangle ,
\end{align}
its diagonal position-space density still contains interference between
different orientations:
\begin{align}
|\Psi_{\rm in}(\mathbf R_X)|^2=
\frac{1}{|N_X|^2}
\int d\Omega_X\,d\Omega_X'\,
\langle\mathbf R_X|\Omega_X\rangle
\langle\Omega_X'|\mathbf R_X\rangle .
\end{align}
Consequently, after the transverse nucleon positions have de-cohered,
the collision geometry continues to depend on the off-diagonal pairs
$(\Omega_X,\Omega_X')$. At the one-body level, this dependence is
encoded in the generalized thickness function
\begin{align}
T_X(\mathbf x;\Omega_X,\Omega_X')
=
\frac{1}{X}
\int dz\,
\frac{
\langle\Omega_X|
\hat\psi_X^\dagger(\mathbf x,z)
\hat\psi_X(\mathbf x,z)
|\Omega_X'\rangle
}{
\langle\Omega_X|\Omega_X'\rangle
}.
\end{align}
Thus, decoherence motivates sampling definite transverse nucleon positions, but it does not reduce the quantum rotor to a classical rotor with a single orientation. The off-diagonal orientation information remains in the decorrelation function and the generalized thickness function.

\twocolumngrid
\bibliographystyle{apsrev4-1}
\bibliography{references}

\end{document}